# An empirical exploration of the diversified R ecosystem

## Author information

Tian-Yuan Huang[1] and Zhilan Lou[1,*]
1 School of Data Science, Zhejiang University of Finance and Economics, Hangzhou, China

First author: Tian-Yuan Huang, E-mail:huangtianyuan@zufe.edu.cn
*Corresponding author: Zhilan Lou, E-mail: zlou@zufe.edu.cn

## Abstract

Born in the late 20s, R is one of the most popular software for statistical computing and graphics. With the development of information technology and the advent of the big data era, great changes have taken place in the R ecosystem. Based on the meta information of the Comprehensive R Archive Network (CRAN) and the bibliometric data of literature citing R, we discovered that while R is initiated by statistics, its development is benefited greatly from computer science and the main user group in academics come from various disciplines such as agricultural science, biological science, environmental science and medical science. In addition, we displayed the collaboration patterns among R developers and analyze the possible effects of collaboration in the R community.

## Introduction

Among the various programming languages, R is famous for its capability in data mining. The TIOBE Index for programming language has found a decreasing trend of R (R Core Team 2021) from 2018 to 2019 (https://www.tiobe.com/tiobe-index/r/). While few people dig into the details of how this index is calculated, this popular index does bring panic and depress to some R users and developers, even those who have been using R for years. But according to the download logs of CRAN (the Comprehensive R Archive Network), neither R software or R packages is facing a reduction in the downloads during this period (Figure 1). On the contrary, there was a remarkable leap in the latter half of 2018. The data shows that August 2018 was the change point of download times (at monthly level), both for R software and R packages. There is no doubt that R has been used by more people and more frequently over time.

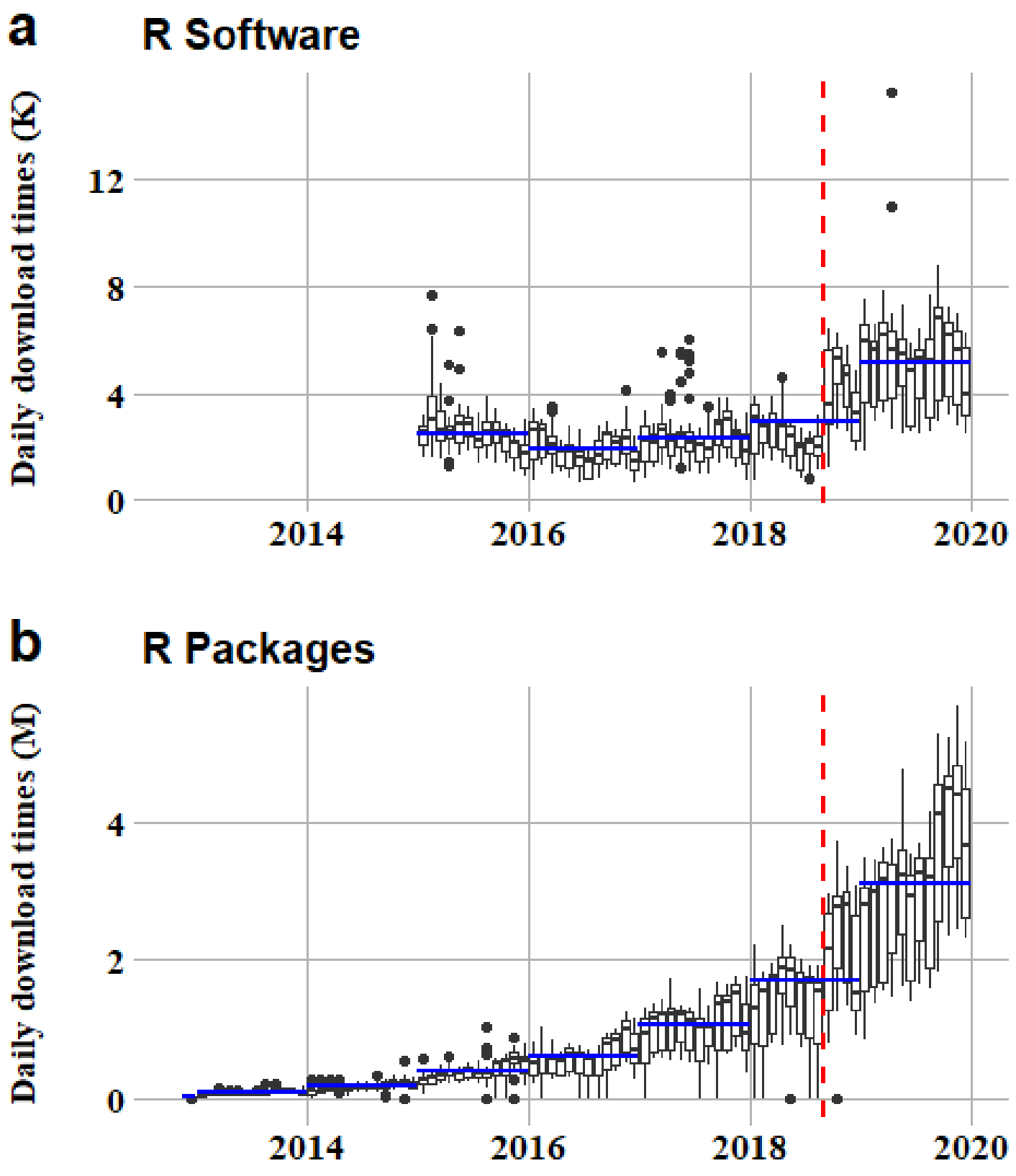

Figure 1: The daily downloads of R software (a) and R packages (b). The boxplots show the daily download times within the month. The blue lines show the average daily download in each year, and the red dashed lines show the change point (Killick and Eckley 2014) of the download. Source of data: http://cran-logs.rstudio.com/.

As a programming language, R's free open-source environment was derived from S language initiated at Bell Labs. The R core was first launched in 1997, and maintained by the R Core Team and R Foundation since then. The public management of R packages is mainly supported by CRAN, other major repositories include BioConductor, R-Forge and GitHub (German et al., 2013; Plakidas et al., 2017). While these repositories have overlapped packages, CRAN remains the center of R package management system with least dependencies (Decan et al., 2015). The usage of R is extensive and gradually it becomes one of the most popular software in scientific research. This leads scientists to pay attentions to the scientific usage of R packages. Related researches include the focus on co-mention network of R packages (Li & Yan, 2018) and citation patterns on specific R package (Li et al., 2019) or general R packages (Li et al., 2017). In addition, numerous studies have showed that researchers and practitioners from various backgrounds are embracing R for its openness and reproducibility (Gentleman et al. 2004, PebesmaNüst and Bivand 2012, Huber et al. 2015, Lowndes et al. 2017, Lai et al. 2019, Kaya et al. 2019).

In this paper, we would like to make an empirical review of R ecosystem based on the ever-improving archived document from CRAN (https://cran.r-project.org/) and the bibliometric database provided by Scopus (https://www.scopus.com/). By ecosystem, we have borrowed the concept from ecology, which highlights that different components of the natural could interact with each other. Here, we try to explore different components (including developers and users) in the R community, and discuss how they might interact with each other. Specifically, we implement our investigation from four aspects: (1) Why are numerous contributors developing R packages? What are these packages providing us? (2) What are the most popular packages in R in the following 15 years since 2005? Why are they so popular? (3) Who are the major users of R in academia? What are the potential trends of R development in scientific community? (4) What is the general collaboration pattern of R community? By exploring these questions and making critical discussions, we intend to gain better understanding of the current R ecosystem and shape the future development of R.

## Methods

To get a comprehensive view on R ecosystem, multiple data sources are collected and utilized in our study. These datasets include: (1) Download information of R and R packages: Download logs of R and R packages could be accessed from Rstudio CRAN Mirror (http://cran-logs.rstudio.com/), and information of daily download times could be retrieved via APIs from 'cranlogs' package (Csárdi 2019); (2) Meta data of R packages on CRAN: The meta data of R packages, such as their authors, maintainers, published date and imported packages, are archived in CRAN. These data could be extracted using 'RWsearch' package (Kiener 2021). The target data in our research was retrieved at January 1st, 2020; (3) Bibliometric data of papers citing R: The bibliometric data could help us explore how R is utilized in academia. Referring to the previous research (Li et al., 2017), our study retrieved the bibliometric data of papers citing R software from Scopus database using advanced search (the advanced query was "REF ( {R: A Language and Environment for Statistical Computing} OR {http://www.r-project.org} ) AND DOCTYPE ( ar )"), yielding 364,928 articles published from 2005 to 2019. The ‘rscopus’ package (Muschelli 2019) was used to facilitate the acquisition of bibliometric data from Scopus database. The bibliometric data contains information of publication year, title, keywords, journal title and ISSN, etc. Scopus Subject Area categories were used to classify the publications into different subjects. The data analyses were carried out in R software (version 4.1.0). The R packages used in the study include ‘tidyverse’ (Wickham et al. 2019), ‘data.table’ (Dowle and Srinivasan 2021), ‘patchwork’ (Pedersen 2020), ‘changepoint’ (Killick and Eckley 2014), ‘lubridate’ (SpinuGrolemund and Wickham 2021), ‘fst’ (Klik 2020), ‘dtplyr’ (Wickham 2021), ‘akc’ (Huang 2020) and ‘tidyfst’ (Huang and Zhao 2020).

## Results

### Initiatives of R development

We extracted the keywords from the “description” field of R packages listed on CRAN and constructed a keyword co-occurrence network to explore the initiatives of developing R packages. To complete this task, we have: (1) Used n-gram tokenizer to segment the corpus. The maximum and minimum of n were 5 and 2 respectively. Unigrams were excluded because they are too granular to carry accurate information. This process yielded the n-grams for each R package. (2) Filtering the n-grams using a user-defined dictionary based on literature keywords. We have used a reverse query to get all the literature that citing R software in Scopus database, then used the author keywords of these literature to form a R-related dictionary. Then the n-grams yielded in the previous step would be filtered by the dictionary (only phrases in the dictionary were retained). (3) Merge synonyms within the R package. Keywords with same stem would be merged into its most frequent form. In addition, if a keyword is a subset of another, they would be merged into the shorter one. For instance, “time series” and “time series analysis would be merged into “time series” (but they would not be merged into “time” because we have excluded unigrams). (4) Construction and visualization of knowledge graph based on keyword co-occurrence in R packages. Nodes in the network is the keywords of the R packages, edge between nodes means the two keywords have co-occurred in the description of the same package. Both the frequency and degree of the keywords were summarised and displayed in the visualization.

In Figure 2, we could find that lots of R packages are data-driven (the degree of “data sets” is the largest). For the top 50 keywords by frequency, only “hypothesis testing” never co-occurs with “data sets”. Most of these keywords fall into the category of statistics, such as “time series”, “linear regression” and “parameter estimation”. This is not surprising, as R was first created by statisticians and designed to be a freely available language and environment for statistical computing and graphics (Ihaka and Gentleman 1996). Nonetheless, there are also other kinds of keywords. For instance, keywords like “user interface” and “web service” should fall into category of computer science, while keyword like “gene expression” and “meta analysis” should be regarded as domain-specific application.

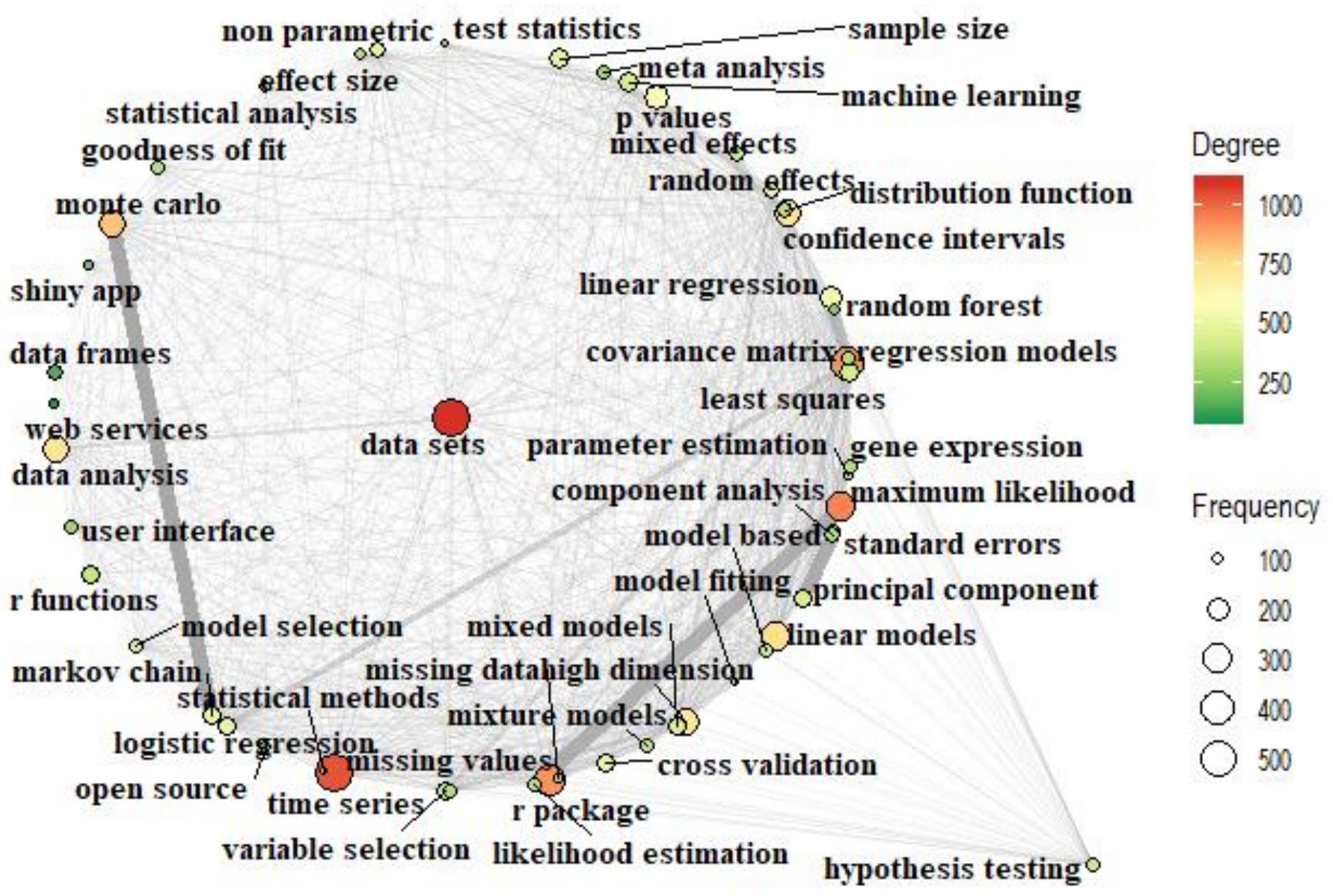

Figure 2: Knowledge graph of R based on package description archived on CRAN. Only top 50 keywords with largest frequency are displayed. The width of edge is proportional to the co-occurrence number. The graph adopted a radial layout around a focal node using 'graphlayouts' package (Schoch 2019), the most frequent keyword is selected as the focal node.

## Popular R packages

Examining the most downloaded R packages in total on CRAN between 2005 and 2019, it could be found that the top 10 R packages (Figure 3a) have a share of ~10% of the total downloads. These packages provide essential functionalities for further development in R ecosystem. Among them, 'ggplot2', 'stringr' (a wrapper of 'stringi'), 'dplyr' might be more well-known to the end-users because they are totally data-oriented and easy to use with consistent and readable APIs. On the other hand, packages like 'Rcpp', 'digest', 'rlang' and 'R6' have brought great convenience to R programmers (especially tool makers), thus usually imported by other packages in R development. Speak of 'magrittr' and 'tibble', their functionalities are considered to be more fundamental. Simply means "and then" in natural language, the operator `%>%` in 'magrittr' is elementary for R as a language. This implementation keeps in line with the logic of human thinking. It reduces development time and improves the readability and maintainability of R codes. As the pipe syntax gaining high recognition and acceptance, in May 18, 2021, a native pipe (`|>`) was introduced in the released R version 4.1.0. On the other hand, considering R as an environment, data frame is an important data structure, and 'tibble' has offered an enhanced class (named as 'tibble' or 'tbl_df') for it. This new class provides end-users to inspect the traditional data frame safely in no time, therefore is adopted by many other R packages and widely used in data science workflows, especially for big data analysis.

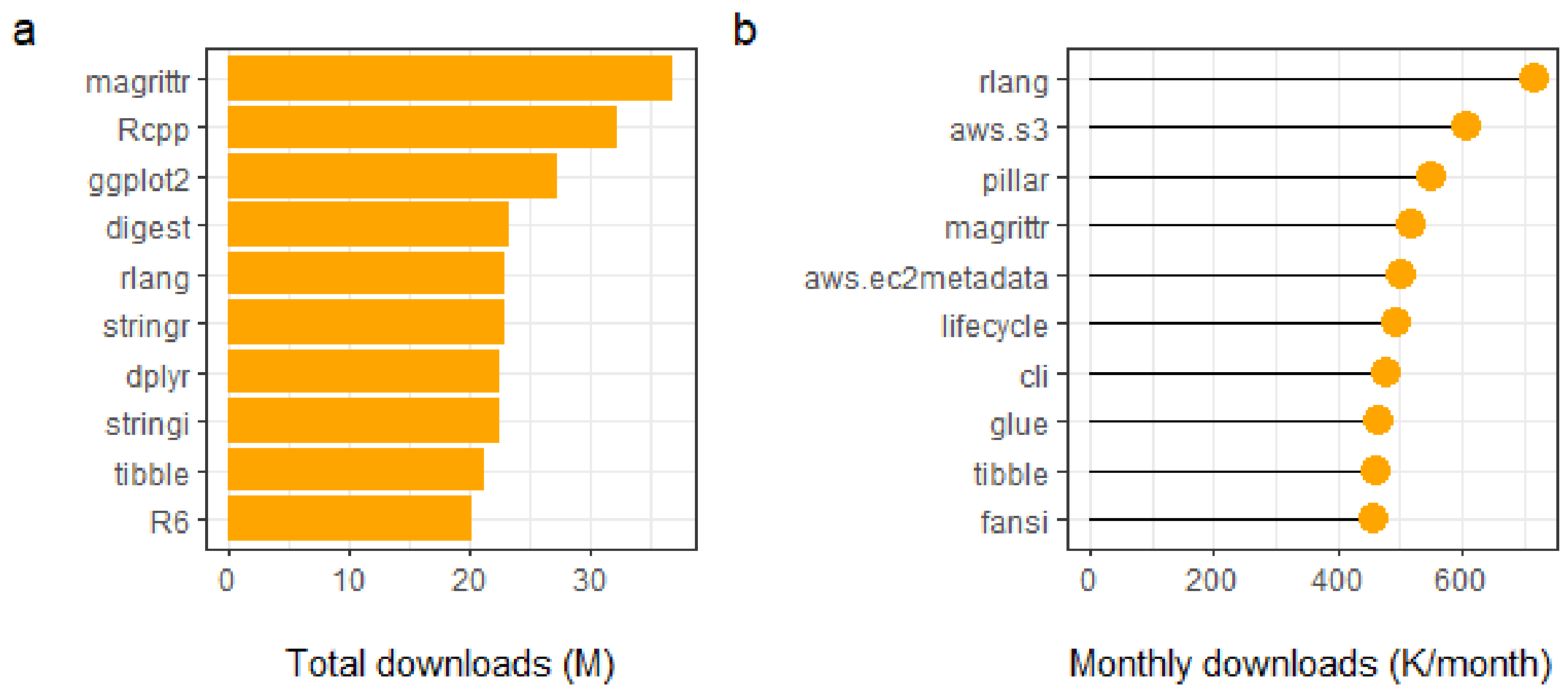

Figure 3: Top 10 R packages by total downloads (a) and monthly downloads (b) from 2005 to 2019.

Turning to the list of monthly downloads (Figure 3b), we could find more packages designed for R developers released recently, including 'pillar', 'lifecycle', 'cli' and 'fansi'. As a complementary component to string operation, 'glue' is also gaining its popularity and become the import package of other utilities (including 'stringr'). Finally, as part of "cloudyr" project (https://cloudyr.github.io/), the rise of 'aws.s3' package and 'aws.ec2metadata' package has revealed another trend: using and developing R in the cloud computing platform. To conclude, the popularity of modern R mainly depends on computer facilities provided by the R environment, which greatly improves the reliability and efficiency of data science workflow.

## Application of R in academia

According to the Scopus database, the number of academic articles citing R software was increasing from 2005 to 2019 (Figure 4a). While only 1,057 were tracked in 2005, the number of articles citing R software has reached 62,168 in 2019. Considering the subject area of these articles, we could find remarkable upward trends of R usage in "Agricultural and Biological Sciences", "Biochemistry, Genetics, and Molecular Biology", "Earth and Planetary Sciences", "Environmental Science", "Immunology and Microbiology", "Medicine", "Psychology" and "Social Sciences" (Figure 4b). During the 15 years' period, "Agricultural and Biological Sciences" and "Environmental Science" turned out to be the most active subject area using R (Figure 4c and Figure 4d). Not only did they published most articles based on R (106,853 and 51,522 respectively), but they also had the highest occupancy of articles utilizing R as well (4.5% and 3.3% respectively). On the other hand, "Computer Science" and "Mathematics" should be considered most relevant to R, but they have relatively low article number and occupation. From 2005 to 2019, there were 10,111 articles categorized as "Mathematics" citing R (0.7% within the subject area), whereas in "Computer Science" there were only 7,164 articles citing R (0.5% within the subject area). This could be attributed to other selection of software associated with "Computer

Science" and "Mathematics", offering scholars extensive options and consequently leading to a reduced proportion of usage for each individual software.

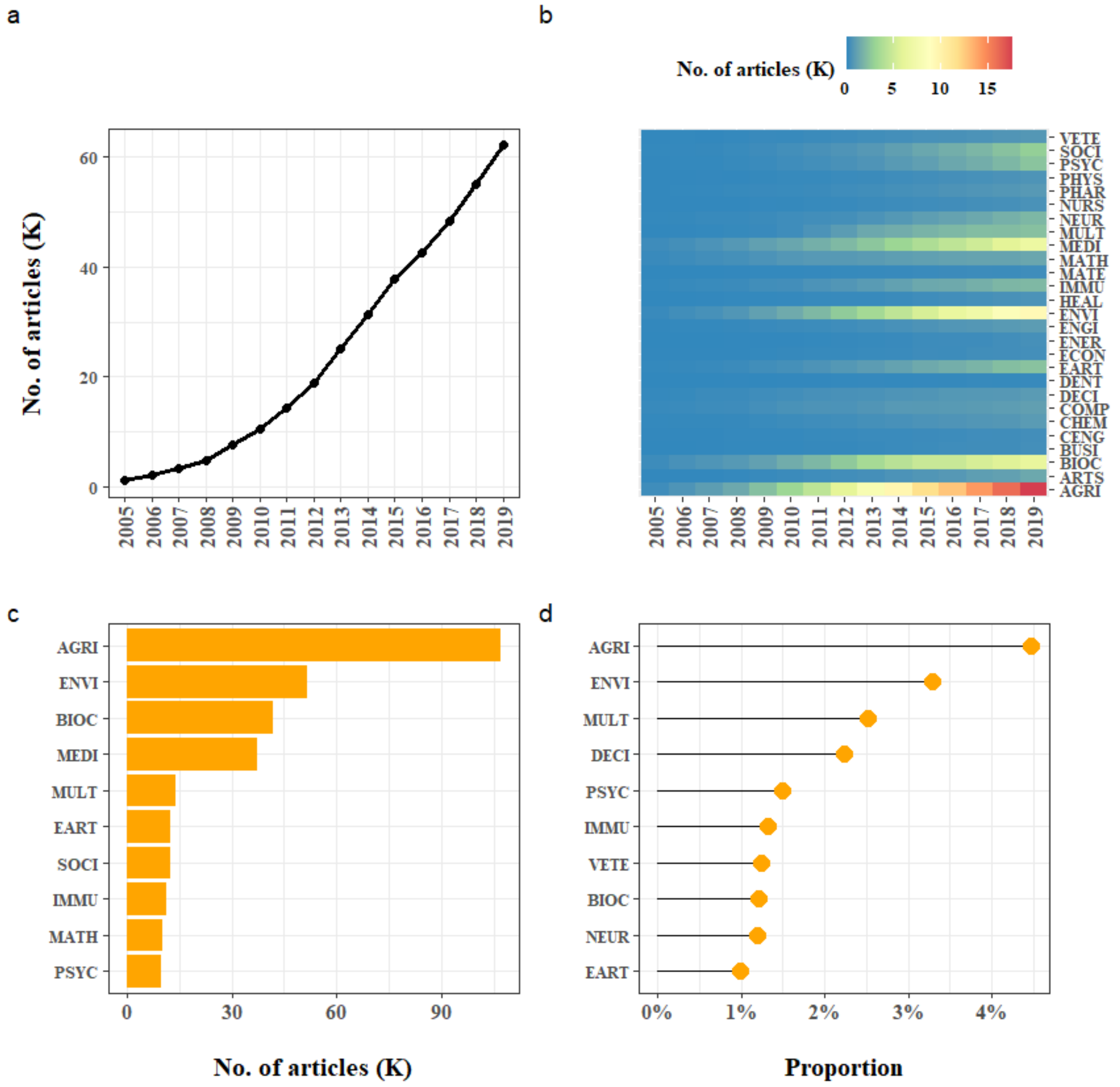

Figure 4: Usage of R in academia between 2005 and 2019. (a) Growth of academic articles citing R software. (b) The number of articles citing R software in different subject areas over time. (c) Top 10 subjects with most articles citing R software. (d) Top 10 subjects with most proportion of articles citing R software. The subject classification is based on the list of Scopus Subject Area (AGRI: Agricultural and Biological Sciences; ARTS: Arts and Humanities; BIOC: Biochemistry, Genetics, and Molecular Biology; BUSI: Business, Management, and accounting; CENG: Chemical Engineering; CHEM: Chemistry; COMP: Computer Science; DECI: Decision Sciences; DENT: Dentistry; EART: Earth and Planetary Sciences; ECON: Economics, Econometrics, and Finance; ENER: Energy; ENGI: Engineering; ENVI: Environmental Science; HEAL: Health Professionals; IMMU: Immunology and Microbiology; MATE: Materials Science; MATH: Mathematics; MEDI: Medicine; MULT: Multidisciplinary; NEUR: Neuroscience; NURS: Nursing; PHAR: Pharmacology, Toxicology, and Pharmaceutics; PHYS: Physics and Astronomy; PSYC: Psychology; SOCI: Social Sciences; VETE: Veterinary).

## Collaboration patterns in R community

An inspection on the meta information of CRAN finds that the package author number follows a long-tail distribution (Figure 5a). By the year of 2019, 6,645 packages are single-authored, followed by 3,677 with two, 2,227 with three, 1,177 with four and 599 with five. For the rest, there are 1,022 packages written by more than five authors. When the package author number is more than one, the package is deemed to be developed via collaboration. From this perspective, we could find more cooperative works in general (8,702 v. 6,645). In examining the potential effects of collaboration (Figure 5b, 5c, 5d), we discovered a positive correlation between the number of package authors and the number of imported packages (Pearson, $r = 0.16$, $p < 2.2e-16$), and updated times per year (Pearson, $r = 0.20$, $p < 2.2e-16$), as well as a weak positive correlation with daily download times (Pearson, $r = 0.09$, $p < 2.2e-16$). This indicates that the collaboration behavior of R developers might possibly help to reuse and integrate more sources in R community and improve development efficiency of R packages. Moreover, the R package coming from team work might gain more popularity as more members are involved. This phenomenon is also common in the academia, where collaborative study attracts more citations than comparable solo research (Wuchty et al., 2007).

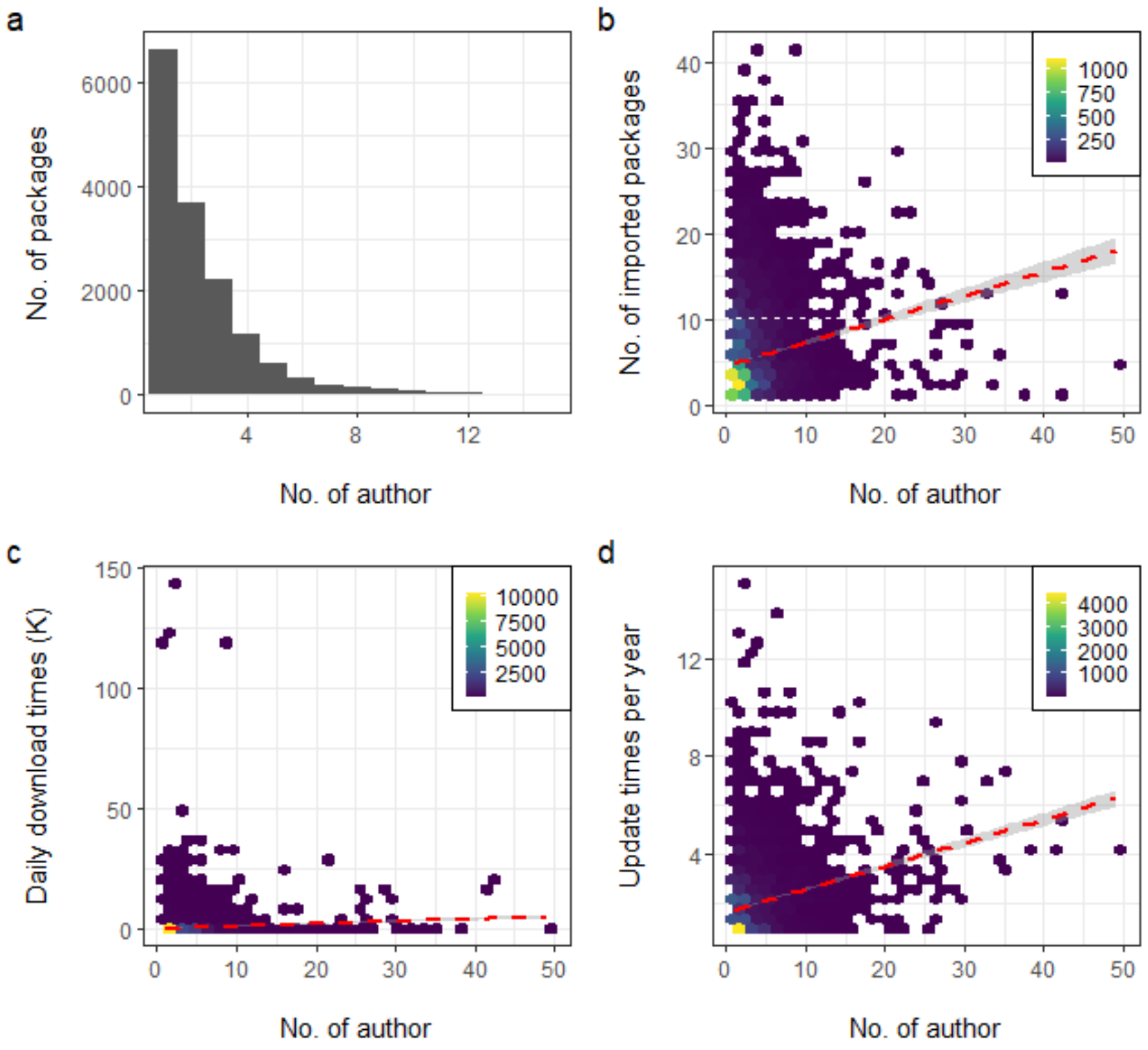

Figure 5: Collaboration patterns in R community. (a) Distribution of R package author number (packages with more than 15 authors are omitted in the visualization). (b) Correlation between author number and number of imported packages. (c) Correlation between author number and daily download times (in the following one month after data collection of package author number). (d) Correlation between author number and update times per year. The legends show how many packages lie in the point area. The red dashed lines show the general linear trends of the points with 95% confidence level in grey.

## Discussion

In history, R was first designed by Ross Ihaka and Robert Gentleman, both statisticians interested in computer programming, as a personal project to build statistical tools in the teaching laboratory (Ihaka 1998). It could be regarded as the fruit of statistics and computer science. But few people would take R as a serious programming language, but rather an environment for statistical computation. Therefore, the development of R has a deep root in statistics, this could also be found in today's R ecosystem (Figure 2). Usually, statistical researchers would propose new ideas and implement these ideas in R, and then tested them in the real world (Figure 6a). A good package designed by statisticians provides the users with easy-to-use APIs and try to guide them to discover more with the additional settings provided by parameters. In this way, even users without any background knowledge could utilize the cutting-edge statistical tools in their work. At the same time, they might discover more issues at the usage and give feedbacks to the package maintainers, which help improve the statistical research in return. It is good practice for developers to keep good documentations for their work. We found that the packages with vignette builder have more downloads than those have not (11,499 v. 7,073 by average in a month). Also, the packages with an URL to provide more information hold higher download records (13,950 v. 3,223 by average per month).

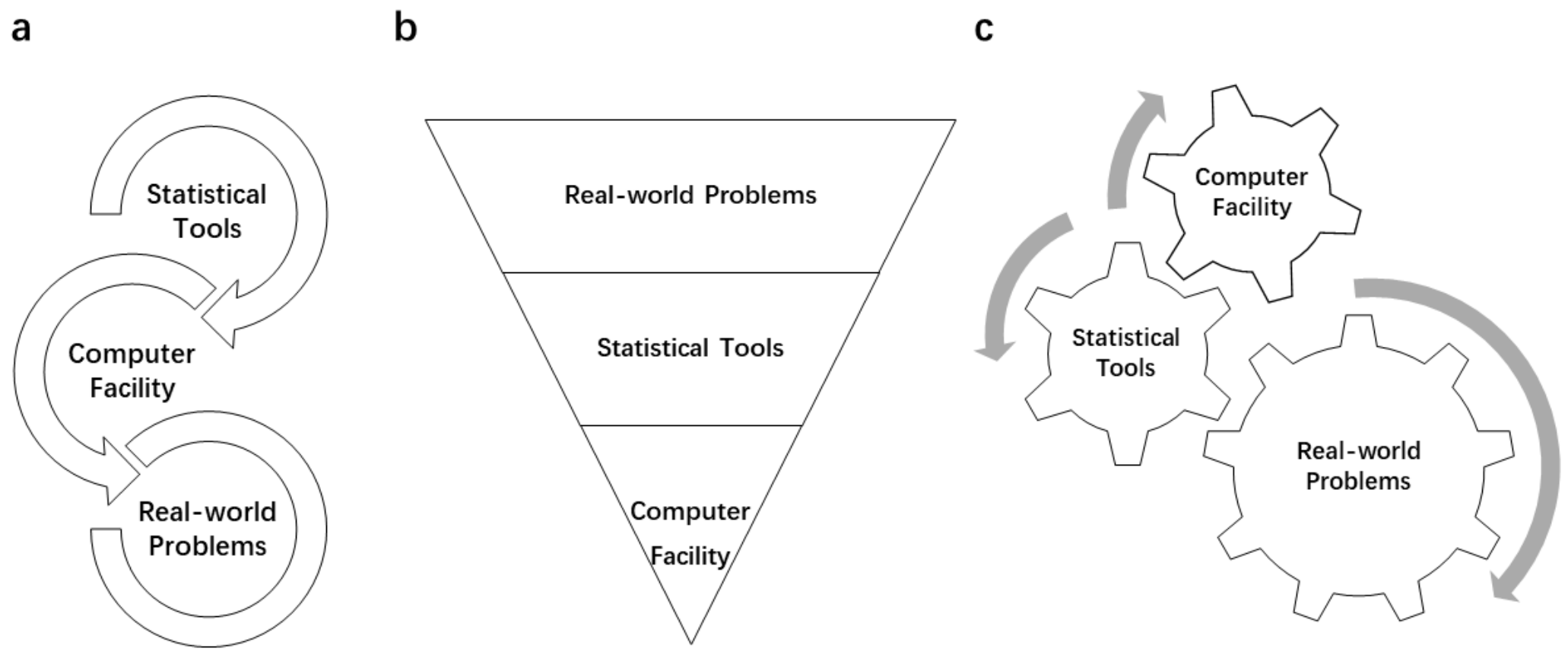

Figure 6: The ecosystem of R from different perspectives. (a) From perspective of statistics. (b) From perspective of computer science. (c) From perspective of practitioners.

While R is initiated by statistics, the recent years have seen lots of popular packages designed for not-so-statistical tasks, such as high performance computing ('Rcpp', 'data.table', 'doParallel', etc.), string operations ('stringi', 'stringr', 'glue', etc.) and connection to other software ('openxlsx', 'pdftools', 'sparklyr', etc.). Generally, these tasks could be categorized as computer facilities, and it is likely that these facilities have acted as footstones in the physical sense (Figure 6b). According to our investigation, the usage of R in computer science is relatively infrequent in academia. As R gets more popular in multiple fields, developers from different backgrounds are considering building a more comprehensive ecosystem for it. The early developers of R might never imagine that one day this light-weighted software could carry out complicated computations on million rows of data within minutes, and at the same time storing and exporting these results to dashboards in expressive graphics and tables. In the field of data science, the versatile R is not in the least inferior to Python, which is well-known for its "glue" feature. On the other hand, while there are lots of statistical tools wrapped in R packages, end-users might find it difficult to use them directly in the real world. The computer facilities, serving as a bridge between statistical tools and real-world problems, could fill this gap. For instance, some advanced statistical methods might be so far only accessible in R, but the real data are stored in different file formats (doc, pdf, tiff, json, etc.). Only by developing the file conversion tools could end-users upscale the power of statistics shared in the community. These tools are so popular that they might gain more attention than other specific statistical tools, because the functionalities they provide are considered to be more general and irreplaceable in the workflow, whereas R provides various alternatives in statistical methods. In the future, the R community should attract more talents with a significant computer science background. The sparks between computer science and statistics would raise more amazing revolutions in the community and lift the ecosystem of R to a higher level.

Although R is rooted in statistics and developed by computer techniques, as the concept and practice of big data swept the world, the fruit of R is shared on a much broader scale. We could find that in academia, a huge group of researchers from various fields are benefiting from R. Usually, they start their research from real-world problems, making hypothesis and seek for the right tools to provide evidence (Figure 6c). One major reason is, these scientific areas are all embracing the advent of big data and moving toward evidence-based qualitative science, and R turns out to be one of the most appropriate tools for this trend. While science emphasizes the academic results should be reproducible, the documented R scripts could provide evidence for inspection and validation in the whole data science workflow. Moreover, as the readability of R is rising rapidly, there seems to be a trend for researchers to communicate and collaborate using R. A well-designed syntax of R language could be comprehended by even non-programmers. In the meantime, these codes could also be run effectively and efficiently in the computer to reproduce the exact same results based on the open shared data. Not only does R provides researchers with powerful computational tools to lower the barrier of statistical implementation, but also provides a big chance to facilitate open science with its wonderful design and community culture. This trend could also create a breeding ground for knowledge transfer and cross-discipline research, as the operational tools (R codes) and statistical logics underneath could be shared and passed from field to field.

With the joint effort from statisticians, computer scientists and practitioners from versatile backgrounds, the ecosystem of R has become unprecedentedly energetic and diverse in the recent two decades. On CRAN we could usually find packages with single authors, because the open source community allow developers to reuse codes freely as long as the license is not violated. Therefore, when they start to develop with R and import functions from other packages, they are standing on the shoulders of each other already. This should be considered an advanced form of communication and collaboration. Nevertheless, our investigation shows that there are positive correlations between package author number and the imports, updates and downloads, which indicates that team work might have improved the diversified creativity, development efficiency and software impact in the R development. Lots of local and international communities of R (rOpenSci, RLadies, RStudio, RUGS, etc.), online or offline, commercial or non-commercial, are emerging and running vigorously these years. While they have various scopes and organizational forms, the core of R remains in every one of them, which is to be free, open and collaborative for a better world. Therefore, just forget about the TIOBE Index or the aggressive comparisons with other programming languages. R is here to stay, and there is a long way to go.

## Declarations

The authors have no conflicts of interest to declare that are relevant to the content of this article. The data and codes of this study are available for reproducibility of the results on request.

## Acknowledgements

This research was supported in part by the National Natural Science Foundation of China, grant number 12001483 (for Lou); in part by the First Class Discipline of Zhejiang – A (Zhejiang University of Finance and Economics – Statistics) (for Lou). The author thanks Liying Yang and Ronald Rousseau for their useful remarks.

## References

Csárdi, G. (2019). cranlogs: Download Logs from the RStudio 'CRAN' Mirror. https://CRAN.R-project.org/package=cranlogs.

Decan, A., Mens, T., Claes, M., & Grosjean, P. (2015, September). On the development and distribution of R packages: An empirical analysis of the R ecosystem. In *Proceedings of the 2015 european conference on software architecture workshops* (pp. 1-6).

Dowle, M. & A. Srinivasan. (2021). data.table: Extension of `data.frame`. https://CRAN.R-project.org/package=data.table.

Download. https://CRAN.R-project.org/package=RWsearch.

Gentleman, R. C., V. J. Carey, D. M. Bates, B. Bolstad, M. Dettling, S. Dudoit, B. Ellis, L. Gautier, Y. Ge, J. Gentry, K. Hornik, T. Hothorn, W. Huber, S. Iacus, R. Irizarry, F. Leisch, C. Li, M. Maechler, A. J. Rossini, G. Sawitzki, C. Smith, G. Smyth, L. Tierney, J. Y. Yang & J. Zhang (2004) Bioconductor: open software development for computational biology and bioinformatics. *Genome Biology,* 5**,** R80.

German, D. M., Adams, B., & Hassan, A. E. (2013, March). The evolution of the R software ecosystem.

In *2013 17th European Conference on Software Maintenance and Reengineering* (pp. 243-252). IEEE.

Huang, T. & B. Zhao (2020) tidyfst: Tidy Verbs for Fast Data Manipulation. *Journal of Open Source Software,* 5**,** 2388.

Huang, T. (2020). akc: Automatic Knowledge Classification. https://github.com/hope-data-science/akc.

Huber, W., V. J. Carey, R. Gentleman, S. Anders, M. Carlson, B. S. Carvalho, H. C. Bravo, S. Davis, L. Gatto, T. Girke, R. Gottardo, F. Hahne, K. D. Hansen, R. A. Irizarry, M. Lawrence, M. I. Love, J. MacDonald, V. Obenchain, A. K. Oleś, H. Pagès, A. Reyes, P. Shannon, G. K. Smyth, D. Tenenbaum, L. Waldron & M. Morgan (2015) Orchestrating high-throughput genomic analysis with Bioconductor. *Nature Methods,* 12**,** 115-121.

Ihaka, R. & R. Gentleman (1996) R: a language for data analysis and graphics. *Journal of computational and graphical statistics,* 5**,** 299-314.

Ihaka, R. (1998) R: Past and future history. *Computing Science and Statistics,* 392-396.

Kaya, E., M. Agca, F. Adiguzel & M. Cetin (2019) Spatial data analysis with R programming for environment. *Human and Ecological Risk Assessment: An International Journal,* 25**,** 1521-1530.

Kiener, P. (2021). RWsearch: Lazy Search in R Packages, Task Views, CRAN, the Web. All-in-One

Killick, R. & I. Eckley (2014) changepoint: An R package for changepoint analysis. *Journal of statistical software,* 58**,** 1-19.

Klik, M. (2020). fst: Lightning Fast Serialization of Data Frames. http://www.fstpackage.org.

Lai, J., C. J. Lortie, R. A. Muenchen, J. Yang & K. Ma (2019) Evaluating the popularity of R in ecology. *Ecosphere,* 10**,** e02567.

Li, K., & Yan, E. (2018). Co-mention network of R packages: Scientific impact and clustering structure. *Journal of Informetrics*, 12(1), 87-100.

Li, K., Chen, P., & Yan, E. (2019). Challenges of measuring software impact through citations: An examination of the lme4 R package. *Journal of Informetrics*, 13(1), 449-461.

Li, K., Yan, E., & Feng, Y. (2017). How is R cited in research outputs? Structure, impacts, and citation standard. *Journal of Informetrics*, 11(4), 989-1002.

Lowndes, J. S. S., B. D. Best, C. Scarborough, J. C. Afflerbach, M. R. Frazier, C. C. O Hara, N. Jiang & B. S. Halpern (2017) Our path to better science in less time using open data science tools. *Nature ecology & evolution,* 1**,** 0160.

Muschelli, J. (2019). rscopus: Scopus Database API Interface. https://CRAN.R-project.org/package=rscopus.

Pebesma, E., D. Nüst & R. Bivand (2012) The R software environment in reproducible geoscientific research. *Eos, Transactions American Geophysical Union,* 93**,** 163-163.

Pedersen, T. L. (2020). patchwork: The Composer of Plots. https://CRAN.R-project.org/package=patchwork.

Plakidas, K., Schall, D., & Zdun, U. (2017). Evolution of the R software ecosystem: Metrics, relationships, and their impact on qualities. *JOURNAL OF SYSTEMS AND SOFTWARE*, 132, 119-146.

R Core Team. (2021). R: A Language and Environment for Statistical Computing. https://www.R-project.org/.

Schoch, D. (2019). graphlayouts: Additional Layout Algorithms for Network Visualizations. https://CRAN.R-project.org/package=graphlayouts.

Spinu, V., G. Grolemund & H. Wickham. (2021). lubridate: Make Dealing with Dates a Little Easier.

https://CRAN.R-project.org/package=lubridate.
Wickham, H. (2021). dtplyr: Data Table Back-End for dplyr. https://github.com/tidyverse/dtplyr.
Wickham, H., M. Averick, J. Bryan, W. Chang, L. D. McGowan, R. François, G. Grolemund, A. Hayes, L. Henry & J. Hester (2019) Welcome to the Tidyverse. *Journal of Open Source Software,* 4**,** 1686.
Wuchty, S., Jones, B. F., & Uzzi, B. (2007). The Increasing Dominance of Teams in Production of Knowledge. *SCIENCE*, 316(5827), 1036-1039. http://doi.org/10.1126/science.1136099